\newif\ifsingle

\newif\ifFullVersion

\ifsingle
\documentclass[11pt,draftclsnofoot, onecolumn]{IEEEtran}		
\else		
\documentclass[10pt,final, twocolumn]{IEEEtran}
\fi

\usepackage{times}
\usepackage{amsmath,dsfont}
\usepackage{amssymb,amsthm}
\usepackage{epsfig,verbatim}
\usepackage{subfigure}
\usepackage{setspace}
\usepackage{color}
\usepackage{cite}
\usepackage{epstopdf}
\usepackage{graphics}
\usepackage{accents}
\usepackage{acronym}
\usepackage[bookmarks,colorlinks]{hyperref}
\usepackage{booktabs}
\usepackage{mathtools}
\usepackage{algorithm}
\usepackage{algorithmicx}
\usepackage{algpseudocode}
\usepackage{amsmath}
\usepackage{enumitem}

\newtheorem{theorem}{Theorem}

\ifsingle
\else
\fi 

\acrodef{adc}[ADC]{analog-to-digital convertor}
\acrodef{cs}[CS]{compressed sensing}
\acrodef{dtft}[DTFT]{discrete-time Fourier transform}
\acrodef{dnn}[DNN]{deep neural network} 
\acrodef{csi}[CSI]{channel state information}
\acrodef{map}[MAP]{maximum a-posteriori probability}
\acrodef{snr}[SNR]{signal-to-noise ratio}
\acrodef{bs}[BS]{base station} 
\acrodef{iot}[IOT]{Interent of Things}
\acrodef{mimo}[MIMO]{multiple-input multiple-output}
\acrodef{mse}[MSE]{mean-squared error}
\acrodef{mmse}[MMSE]{minimum mean-squared error}
\acrodef{pdf}[PDF]{probability density function}
\acrodef{rv}[RV]{random variable}
\acrodef{fec}[FEC]{forward error correction}
\acrodef{dma}[DMA]{dynamic metasurface antenna}
\acrodef{lti}[LTI]{linear time-invariant}
\acrodef{wss}[WSS]{wide-sense stationary}
\acrodef{psd}[PSD]{power spectral density}
\acrodef{ser}[SER]{symbol error rate} 
\acrodef{ber}[BER]{bit error rate} 
\acrodef{sgd}[SGD]{stochastic gradient descent} 
\acrodef{isi}[ISI]{intersymbol interference}  
\acrodef{awgn}[AWGN]{additive white Gaussian noise} 
\acrodef{ut}[UT]{user terminal} 
\acrodef{mmw}[mmWave]{millimeter wave}

\IEEEoverridecommandlockouts

\title{ 

Beamforming Design for Integrated Sensing and Wireless Power Transfer Systems}
\author{\IEEEauthorblockN{Qianyu Yang,~\IEEEmembership{Student Member,~IEEE},  Haiyang Zhang,~\IEEEmembership{Member,~IEEE}, and Baoyun Wang,~\IEEEmembership{Senior Member,~IEEE}  \\
	}   
	
	\thanks{ 
		 
		}

}

\begin{document}
	
	\maketitle
	\pagestyle{empty}
	\thispagestyle{empty}

\begin{abstract} 
This letter proposes a new concept of integrated sensing and wireless power transfer (ISWPT), where radar sensing and wireless power transfer functions are integrated into one hardware platform. ISWPT provides several benefits from the integrating operation such as system size, hardware cost, power consumption, and spectrum saving, which is envisioned to facilitate future 6G wireless networks. As the initial study, we aim to characterizing the fundamental trade-off between radar sensing and wireless power transfer, by optimizing transmit beamforming vectors. We first propose a semi-definite relaxation-based approach to solve the corresponding optimization problem globally optimal, and then provide a low-complexity sub-optimal solution. Finally, numerical results verify the effectiveness of our proposed solutions, and also show the trade-off between radar sensing and wireless power transfer.

	\end{abstract}

	\section{Introduction}

The capability of Internet of Everything (IoE) services is expected to be an important evolution metric of Beyond-5G networks \cite{5G}, for supporting many new emerging applications such as intelligent vehicle networks \cite{IoV}. These ambitious targets rely on sensing ability of wireless netwoks with radar sensing \cite{Sensing} and massive low-power sensors \cite{low}. For example, intelligent vehicle networks that require sensors to know the road and vehicle status through radar beams emitted by roadside units \cite{V2X}. Meanwhile, due to the size and hardware cost constraints, these low-power sensors are usually powered by a limited battery or even be battery-less \cite{energy-save,WPT3}.  Radio frequency-based wireless power transfer (WPT) is regarded as a promising technology to solve the energy shortage problem of low-power devices, which is capable of charging wireless devices in a wireless manner \cite{WPT,WPT2,SWIPT}. 

It is worth pointing out that conventional radar sensing and WPT are studied separately as in the works mentioned above.
In this letter, we propose a novel concept that integrates radar sensing and WPT into one hardware platform, and call the considered scenario as integrated sensing and wireless power transfer (ISWPT). 
The proposed ISWPT system has many promising application scenarios such as intelligent vehicle networks \cite{V2E}, enabling the implementation of radar sensing and charge sensors simultaneously.
With such integrated operation, ISWPT has several advantages compared to conventional separately designs: i) the integration conserves the device cost and spectrum resources, which is significant to future 6G networks; ii)  the radar beam usually provides continuous coverage of a range of areas, which facilitates a stable energy supply to equipment in the area, and meanwhile, the collecting energy from radar signals also reduces power waste of radars.

We consider a novel ISWPT scenario, where both the radar sensing and WPT are integrated into one hardware platform. The multi-antenna transmitter generates one type of waveforms to achieve the two functions simultaneously. As these waveforms share the same antenna and transmit power resources, there exists a fundamental performance trade-off between radar sensing and WPT. We characterize this fundamental performance trade-off by optimizing the transmitted signal beamforming vectors for the ISWPT system. The main contributions of this letter are summarized as follows. 
\begin{itemize}
\item To the best of our knowledge, we are the first to propose the concept of ISWPT, which enables the radar sensing and wireless power transfer sharing the same hardware platform, spectrum and power resources. Specifically, we consider a general ISWPT system, where a multiple-antenna transmitter serves multiple energy receivers and senses an area simultaneously. 
\item Under the considered scenario, we characterize the fundamental performance trade-off between radar sensing and WPT, by optimizing the transmit beamforming vectors. Though the corresponding beamforming design problem is non-convex, we solve it globally optimally by utilizing the SDR approach. Moreover, we also propose a low-complexity sub-optimal beamforming design to further reduce the complexity of the optimal design.
\item We provide numerical results to demonstrate the fundamental performance trade-off between radar sensing and WPT. Numerical results also verify the effectiveness of the proposed beamforming designs for the novel ISWPT system.
\end{itemize}

\textit{Notation:} Scalar variables, vectors and matrices are represented with lower letters, lower bold letters, and capital bold letters, respectively (e.g., $x$, $\bf x$, and $\bf X$, respectively). The term $\mathbb{C}^{N \times N}$ denotes a complex space of dimension ${N \times N}$. $\left( \cdot \right)^H $ denotes the Hermittian operators. For a matrix $\bf R$, ${\bf R} \succeq  {\bf 0}$ means that $\bf R$ is positive semidefinite, while ${\bf R}_{\left[ i:j,i:j \right]}$ denotes the extraction operation for corresponding rows and columns of $\bf R$. ${\bf 0}_{M \times N}$ denotes the $M \times N$ dimension zero matrix.



\section{System Model}
\label{sec:Model}


\subsection{Transmitted Signal Model}
\label{subsec:SignalModel}

\begin{figure}[t!]
\centering
\includegraphics[scale=0.3]{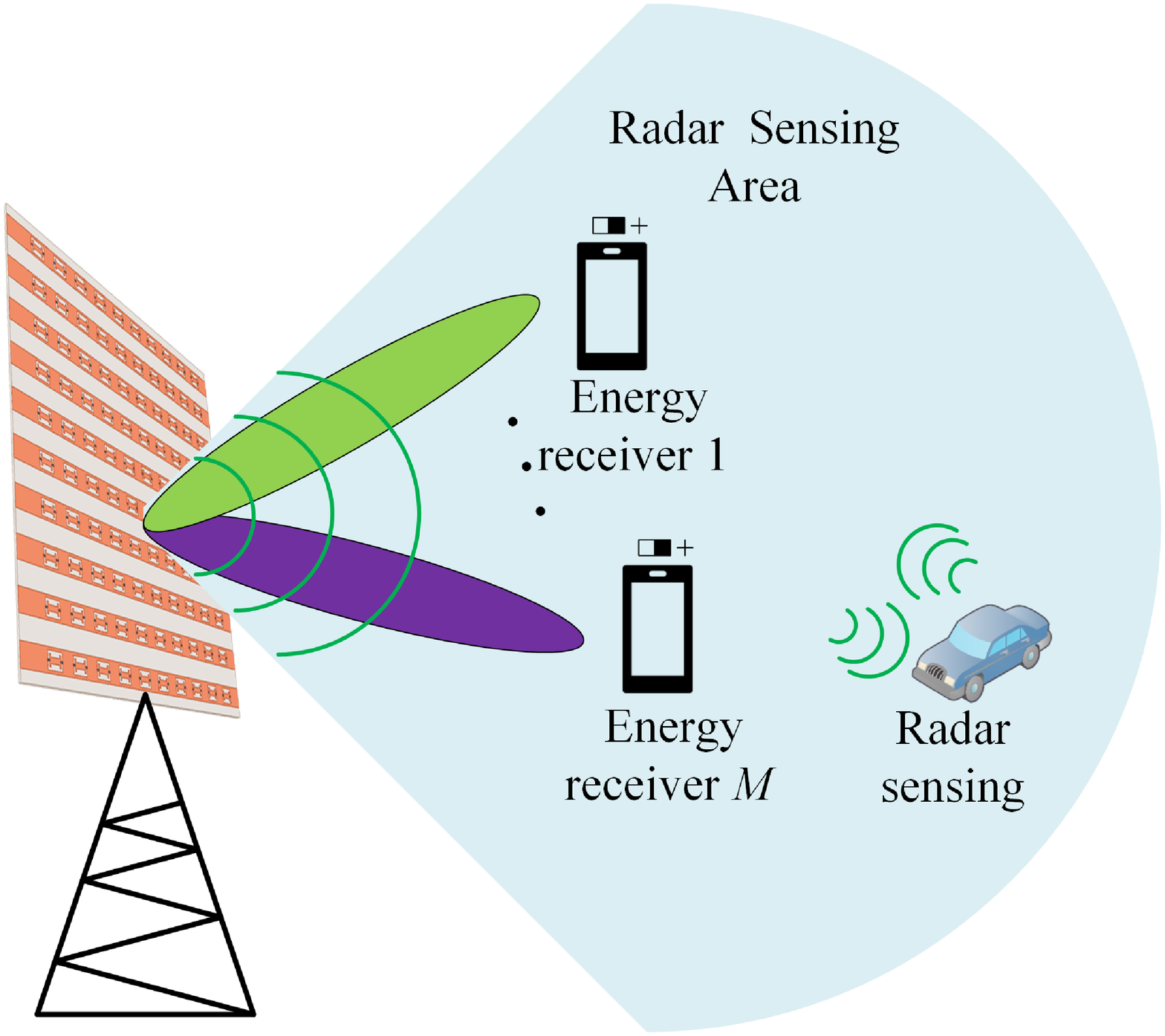}
\caption{Illustration of one ISWPT scenario, where one multi-antenna array simultaneously senses a range of area and charges multiple energy receivers.
} 
\label{fig1}
\end{figure}

We consider a novel ISWPT system, where one transmitter equipped with $N$ antennas serves $M$ energy receivers and sense an area simultaneously, as shown in Fig.~\ref{fig1}.
Unlike conventional integrated radar and communication setups where both the radar and information signals are required, here we consider utilizing only radar signals to realize sensing and WPT. This is because energy receivers can also harvest energy from radar signals, and there is no need to distinguish radar signals and WPT signals in our considered ISWPT systems. 
Thus, the transmitted base-band signal can be expressed as
\begin{equation}
\label{beam}
{\bf x} = \sum_{n=1}^{N} {\bf w}_{n} s_n, 
\end{equation}
where ${\bf w}_{n} \in \mathbb{C}^{N \times 1}$, denotes the beamforming vector for radar sensing and energy receivers; $ s_n $ denotes the transmitted radar symbols, which are independently and uncorrelated with each other and satisfy $\mathbb{E}\left[|s_n|^2\right]=1$. 

Define the covariance matrix of the transmitted signal by ${\bf R}=\mathbb{E}\left[{\bf x} {\bf x}^{H}\right]$, according to \eqref{beam}, which is given by
\begin{equation}
\label{corv}
{\bf R} = \sum_{n=1}^{N} {\bf w}_{n} {\bf w}_{n}^H .
\end{equation}

Consider the per-antenna transmit power constraints, the transmit covariance matrix  ${\bf R}$ should satisfy
\begin{equation}
\label{powerlimit}
 {\bf R}_{n,n} =  P_{\rm t}/ N,
\end{equation}
where $P_{\rm t}$ denotes the total transmit power budget.

\subsection{Radar Sensing and WPT Metrics}
\label{subsec:Radar WPT}

For radar sensing, the baseband signal at direction $\theta$ is expressed as $y \left[ \theta \right]={\bf a}^{H} \left( \theta \right) {\bf x}$, where ${\bf a} \left( \theta \right) = \left[ 1, e^{j2\pi \Delta \theta},\cdots,e^{j2\pi({\rm N-1}) \Delta \theta} \right]^T \in \mathbb{C}^{N \times 1}$ is the array steering vector of direction $\theta$ with $\Delta$ being the spacing between adjacent antenna elements. 
The goal of radar beamforming design is to match the desired beam pattern. 
Consequently, similar to \cite{radar}, we can use the following loss function to evaluate the radar performance
\begin{equation}
\label{perf_r}
\begin{aligned}
L_{\rm r} \left({\bf R} \right) & = \frac{1}{L} \sum_{l=1}^{L}\left| \alpha d\left(\theta_{l}\right) - {\bf a}^{H} \left( \theta_l \right) {\bf R} {\bf a} \left( \theta_l \right)\right|^{2}, 
\end{aligned}
\end{equation}
where $d\left(\theta \right)$ is the given desired beam pattern, and $\left\{ \theta_l \right\}_{l=1}^{L}$ denote $L$ sampled angle grids, and $\alpha$ is a scaling factor.

For WPT operation, the received power of the $m$-th user can be represented as \cite{SWIPT2}
\begin{equation}
\label{pe}
\begin{aligned}
&P_m \left({\bf R}\right) =\zeta {{\bf g}_m} {\bf R} {\bf g}^H_m,
\end{aligned}
\end{equation}
where $0<\zeta<1$ is a constant energy conversion efficiency, and ${\bf g}_m \in \mathbb{C}^{1 \times N}$ denotes the wireless channel between  the transmitter and the $m$-th energy receiver. 


To unify units with radar metric, the performance metric of WPT in the ISWPT system is given by
\begin{equation}
\label{per_e2}
\begin{aligned}
L_{\rm e} \left({\bf R} \right) = \frac{1}{ M} \sum_{m=1}^{ M}\left| {P_m^{*}} - P_m \left( {\bf R} \right) \right|^{2}, 
\end{aligned}
\end{equation}
where $\{P_m^{*}\}_{m=1}^M$ denotes the desired received power of each energy receiver. Without loss of generality, we set $N \geq M$. Particularly, $\{P_m^{*}\}_{m=1}^M$ are  received power of each energy receiver, associated with the optimal solution of the following sum-received power maximization problem 
\begin{equation}
\label{perf_e}
\begin{aligned}
\max _{{\bf R} \succeq  {\bf 0}}~ \sum_{m=1}^{ M} P_m \left( {\bf R } \right) \quad s.t.~~ \eqref{powerlimit}.
\end{aligned}
\end{equation}

Problem \eqref{perf_e} is convex that can be directly solved using existing convex optimization tools such as CVX \cite{grant2014cvx}.



\subsection{Problem Formulation}
\label{subsec:Beamforming}

From \eqref{perf_r} and \eqref{per_e2}, we know that both $L_{\rm r} \left({\bf R} \right)$ and $L_{\rm e} \left({\bf R} \right)$ are functions of the transmit covariance matrix $\bf R$. In general, the optimal $\bf R$ that minimize the radar metric $L_{\rm r} \left({\bf R} \right)$ is not optimal for WPT metric $L_{\rm e} \left({\bf R} \right)$, and vice versa. Therefore, there exists a fundamental trade-off between the radar sensing and WPT performance in ISWPT systems. In this letter, we would like to characterize this fundamental performance trade-off between radar sensing and power transfer by optimizing the transmitted beamforming vectors (or equivalently transmit covariance matrix). Mathematically, our interested beamforming design problem is formulated as 
\begin{equation}
\label{problem1}
\begin{aligned}
\min _{\{ {\bf w}_n\}} ~&\left( 1-\rho \right) L_{\rm r} \left({\bf R}\right) + \rho L_{\rm e} \left({\bf R} \right), \\  s.t.~~ & {\bf R} = \sum_{n=1}^{N} {\bf w}_{n} {\bf w}_{n}^H, \\ &{\bf R}_{n,n} =  P_{\rm t}/ N, \\ &{\bf R} \succeq  {\bf 0},
\end{aligned}
\end{equation}
where $0 \leq \rho \leq 1$ is a weighting factor that denotes the performance priority for radar and power transfer tasks in the ISWPT system. The extreme case, e.g.,  $\rho=0$ or  $\rho=1$ corresponds to the scenario of only radar or only WPT.



\section{ Beamforming Design }
\label{sec:solution}


\subsection{Optimal Beamforming Design} 
\label{subsec:SDR}

The optimization problem \eqref{problem1} is not convex because of the quadratic equality constraint, and thus it is difficult to solve. 
To deal with this problem, we use the semidefinite relaxation (SDR) approach \cite{luo2010semidefinite} to transform \eqref{problem1} into a convex semidefinite programming (SDP) problem. For this purpose, we define
\begin{equation}
\label{Ri}
{\bf R}_{i}= {\bf w}_{i} {\bf w}_{i}^{H} ~~ i= 1, \cdots,  N.
\end{equation}

By substituting \eqref{Ri} into \eqref{problem1}, we can rewritten it as an equivalent quadratic semidefinite programming (QSDP) with rank-one constraints

\begin{subequations}
\label{problem2}
\begin{align}
\min _{\{ {\bf R}_{i}\}} & \label{10a} ~\left( 1-\rho \right) L_{\rm r} \left({\bf R}\right) + \rho L_{\rm e} \left({\bf R} \right),\\ \text { s.t. } & {\bf R }= \sum_{i=1}^{ N} {\bf R}_{i}, \\ &{\bf R}_{n,n} =  P_{\rm t}/ N, \\ \label{10d} &{\bf R}_i \succeq  {\bf 0}, 
\text {rank}\left( {\bf R}_{i} \right)=1.
\end{align}
\end{subequations}

In order to make \eqref{problem2} a convex problem, we adopt the SDR approach to omit the non-convex rank-1 constraint in \eqref{10d}. The resulting problem is convex which can be solved directly using existing convex optimization solvers such as CVX.

Let $\{ {\bf \hat{R}}_{i}\}$ denote the optimal solution to the relaxed problem of \eqref{problem2}. We note that  $\{ {\bf \hat{R}}_{i}\}$ may not satisfy the rank-1 constraint due to the relaxation operation. To deal with this problem, commonly used approaches are to find a rank-1 approximation solution by using Randomization procedure \cite{luo2010semidefinite}, which will inevitably degrade the performance. 
Instead of using these approximation methods, we propose an efficient approach that is capable of directly constructing the optimal solution of \eqref{problem2} from the relaxed solution, which is summarized as the following theorem.
\begin{theorem}
 \label{thm:MSE}
Define $\{ {\bf \hat{R}}_{i}\}$ as the optimal solution of the relaxed problem of \eqref{problem2}. Then the rank-1 optimal solution of \eqref{problem2}, denoted by ${\bf \widetilde{R}}_{i}$, can be constructed by
\begin{equation}
\label{tight_Ri}
{\bf \widetilde{R}}_{i}= {\bf \widetilde{w}}_i{\bf \widetilde{w}}_i^{H}, ~ i= 1, \cdots, N,
\end{equation}
where ${\bf \widetilde{w}}_i$ is the $i$-th column vector of the matrix ${\bf \widetilde{W}} = {\bf \widetilde{U}} {\widetilde{\boldsymbol \Lambda}}^{\frac{1}{2}} {\bf \widetilde{U}}^{H}$, with ${\bf \widetilde{U}}$ and $\widetilde{\boldsymbol \Lambda}$ denoting the eigenvector matrix and the eigenvalue diagonal matrix of ${\bf \hat{R}} = \sum_{i=1}^{ N} {\bf \hat{R}}_i$, respectively.
\end{theorem}
\begin{IEEEproof} 
Define ${\bf \hat{R}} = \sum_{i=1}^{ N} {\bf \hat{R}}_i$. ${\bf \hat{R}}$ is a $N$-by-$N$ positive semi-definite matrix, and its eigenvalue decomposition can be expressed as
\begin{equation}\label{eq:EVD}
{\bf \hat{R}} = {\bf \widetilde{U}}{\widetilde{\boldsymbol \Lambda}} {\bf \widetilde{U}}^{H},
\end{equation}
where ${\bf \widetilde{U}}$ and ${\widetilde{\boldsymbol \Lambda}}$ are the eigenvector matrix and the eigenvalue diagonal matrix of ${\bf \hat{R}}$, respectively. 

Define ${\bf \widetilde{W}} = {\bf \widetilde{U}} {\widetilde{\boldsymbol \Lambda}}^{\frac{1}{2}} {\bf \widetilde{U}}^H$ and ${\bf \widetilde{W}}=[{\bf \widetilde{w}}_1, \cdots, {\bf \widetilde{w}}_N]$. Since ${\bf \widetilde{U}}$ is a unitary matrix, we have
\begin{equation}\label{eq:U}
{\bf \hat{R}} = {\bf \widetilde{W}}{\bf \widetilde{W}}^{H}= \sum_{i=1}^N {\bf \widetilde{w}}_i{\bf \widetilde{w}}_i^H.
\end{equation}

Define a new set of rank-1 and positive semi-definite matrices, $\left\{{\bf \widetilde{R}}_{i}\right\}$, given by
\begin{equation}
\label{eq:rank-1}
{\bf \widetilde{R}}_{i}= {\bf \widetilde{w}}_i{\bf \widetilde{w}}_i^{H}, ~ i= 1, \cdots, N.
\end{equation}

From \eqref{eq:U} and \eqref{eq:rank-1}, we can verify that ${\bf \hat{R}} = \sum_{i=1} {\bf \widetilde{R}}_{i}$. This implies that $\left\{{\bf \widetilde{R}}_{i}\right\}$ can achieve the same objective function as the SDR optimal solution. Meanwhile, $\left\{{\bf \widetilde{R}}_{i}\right\}$ can satisfy all the constraints in \eqref{problem2}. Therefore, $\left\{{\bf \widetilde{R}}_{i}\right\}$ are the globally optimal solution to \eqref{problem2}.

\end{IEEEproof}

According to Theorem \ref{thm:MSE}, the corresponding optimal beamforming vectors $\left\{{\bf w}_{i}\right\}$ to \eqref{problem1} can be recovered from ${\bf \widetilde{R}}_{i}$ directly.

The worst case complexity to solve the QSDP \eqref{problem2} is $\mathcal{O} \left( N^{6.5}N^{6.5}  \log (1 / \epsilon) \right)$ with the primal-dual interior-point algorithm \cite{complex}, where $\epsilon$ is the solution accuracy. 



\subsection{
Sub-optimal Beamforming Design
}
\label{subsec:Sub}
In order to reduce the high complexity of optimal beamforming design from SDR method, we further propose a sub-optimal beamforming design. Specifically, we still adopt radar beamforming vectors to implement radar sensing and WPT simultaneously, but consider the radar beamforming matrix has the following structure: 
\begin{equation}
\label{eq:W_structure}
{\bf W}=\left[ {\bf W}_{\rm MRT}, {\bf W}^{\bot}_{\rm MRT} \right]
\end{equation}
where ${\bf W} \in \mathbb{C}^{N \times N}$ denotes the beamforming matrix; ${\bf W}_{\rm MRT}\in \mathbb{C}^{N \times M}$ is the maximum ratio transmission (MRT) beams towards energy receivers, i.e.,  ${\bf W}_{\rm MRT}={\bf G}^H{\bf \Lambda}$, where ${\bf G}^H = \left[ {\bf g}^H_1, \cdots, {\bf g}^H_M \right]$ and ${\bf \Lambda} \in \mathbb{C}^{M \times M}$ is a diagonal matrix with diagonal element $\lambda_m$ denoting the transmit power allocated for the corresponding beamforming vector; ${\bf W}^{\bot}_{\rm  MRT}$ is the orthogonal complement matrix of ${\bf W}_{\rm MRT}$.

By substituting \eqref{eq:W_structure} into problem \eqref{problem1}, we then have 
\begin{subequations}
\label{problem2_power}
\begin{align}
\min _{ {\bf W}} ~~& \left( 1-\rho \right) L_{\rm r} \left({\bf R}\right) + \rho L_{\rm e} \left({\bf R} \right), \\ 
s.t.~~ \label{15b} & {\bf R}= {\bf W}{\bf W}^H, \\ &{\bf R}_{n,n} =  P_{\rm t}/ N, {\bf R} \succeq  {\bf 0}, \\
& \label{15d} {\bf G} {\bf W} = \left[ {\bf G}{\bf G}^{H}{\bf \Lambda}, {\bf 0}_{M \times (N-M)} \right], 
\end{align}
\end{subequations}
where the \eqref{15d} comes from the fact that  ${\bf G} {\bf W}^{\bot}_{\rm MRT}  = {\bf 0}$.

Though problem \eqref{problem2_power} is non-convex due to the quadratic equality constraint, we can solve it globally optimal. To this end, we need the following theorem. 
\begin{theorem}
 \label{thm:MRT}
There always exists the covariance matrix ${\bf R} \in \mathbb{C}^{N\times N}$ that satisfies
\begin{equation}
\label{MRTR}
{\bf G} {\bf R } {\bf G}^{H} = {\bf G}{\bf G}^{H}{\bf \Lambda}^2 {\bf G}{\bf G}^{H}, 
\end{equation}
and for any ${\bf R}$ satisfies \eqref{MRTR}, there exists ${\bf W} \in \mathbb{C}^{N \times N}$ that satisfies \eqref{15d} with ${\bf R}= {\bf W}{\bf W}^H$ and has the structure
\begin{equation}
\label{MRT W}
{\bf W} = {\bf DU}^H{\bf U}_{\rm H},
\end{equation}
where $\bf D$ is the Cholesky decomposition of $\bf R$, $\bf U$ is the unitary matrix obtained from the the QR decomposition of ${\bf G}{\bf D}$, ${\bf U}_{\rm H}$ is the unitary matrix obtained from the the QR decomposition of $\left[ {\bf G}{\bf G}^{H}{\bf \Lambda},{\bf 0}_{M \times (N-M)} \right]$.

\end{theorem}

\begin{IEEEproof} Please refer to Appendix \ref{Appendix:B}.

\end{IEEEproof} 
From Theorem ~\ref{thm:MRT}, we know the constraint \eqref{15d} is equivalent to \eqref{MRTR}, and once $\bf R$ is determined, $\bf W$ can be recovered from \eqref{MRT W} directly. Thus, the optimal solution to \eqref{problem2_power}  can be obtained by solving the problem below
\begin{subequations}
\label{problem3}
\begin{align}
\min _{\bf R, \Lambda} \quad &  \left( 1-\rho \right) L_{\rm r} \left({\bf R}\right) + \rho L_{\rm e} \left({\bf R} \right),\\  s.t. ~~ & {\bf G} {\bf R } {\bf G}^{H} = {\bf G}{\bf G}^{H}{\bf \Lambda}^2 {\bf G}{\bf G}^{H}, \\ & \label{12c}{\bf R}_{n,n} =  P_{\rm t}/ N, \\ &{\bf R} \succeq  {\bf 0}.
\end{align}
\end{subequations} 


Problem \eqref{problem3} is a convex QSDP of $\bf R$. The worst case complexity of solving \eqref{problem3} is $\mathcal{O} \left( N^{6.5} \log (1 / \epsilon) \right)$, which is much lower than that of the optimal beamforming design.

%

	\section{Numerical Evaluations}
	\label{sec:Sims}

In this section, we provide numerical results to demonstrate the fundamental trade-off between radar sensing and WPT under an novel ISWPT scenario. 
In our experiments, we set $P_t=1$~W, $N=10$, $M = 3$, and $\zeta=0.5$. The wireless channels are generated according to the standard Complex Gaussian distribution. The channel attenuation from the transmitter to energy receivers is identical and set to be $30$~dB. We assume the ideal radar beam pattern consisting of three main beams with directions $\left[-40^\circ, 0^\circ, 40^\circ \right]$,  and the direction grids in \eqref{perf_r} are set as a uniform sample in the range of $-90^\circ$ to $90^\circ$ with $0.1^\circ$ interval. 

\begin{figure}
\centering
\includegraphics[scale=0.5]{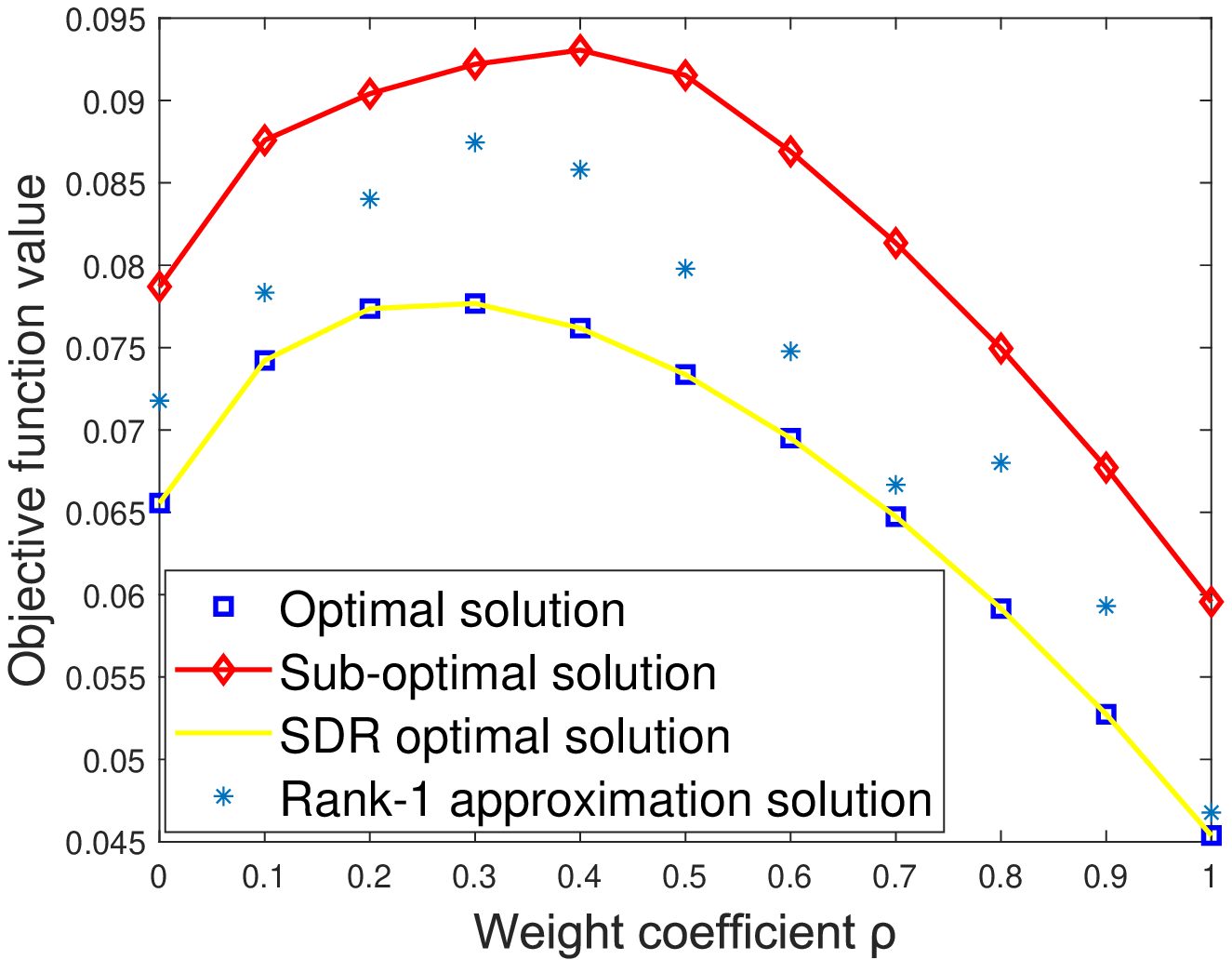}
\caption{The objective function value of  \eqref{problem2} versus different weight coefficient. } 
\label{fig4}
\end{figure}

In Fig.~\ref{fig4}, we compare the performance of our proposed schemes in terms of minimizing the objective function value of problem \eqref{problem2} with two benchmark schemes (SDR optimal solution and Rank-1 approximation solution \cite{luo2010semidefinite}), under different values of weight coefficient $\rho$. 
From Fig.~\ref{fig4}, we can see that our proposed optimal solution achieves the same performance as the SDR optimal solution, which implies that our proposed optimal solution is the globally optimal solution to problem \eqref{problem2}. Moreover, as expected, our proposed optimal solution performs better than the Rank-1 approximation solution and the proposed low-complexity sub-optimal solution.

\begin{figure} 
  \centering 
  \subfigure[$\rho=0.1$]{ 
    \label{fig:subfig:near-field}
    \includegraphics[width=3in]{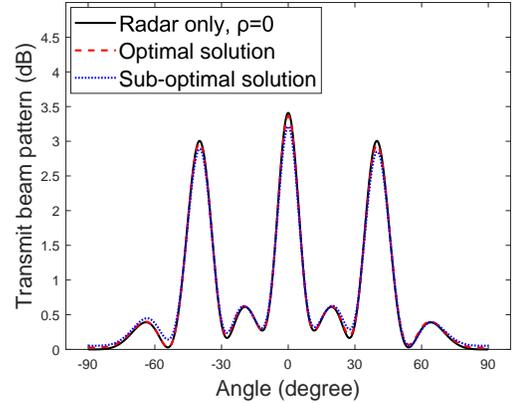} 
  } 
  \subfigure[$\rho=0.5$]{ 
    \label{fig:subfig:far-field} 
    \includegraphics[width=3in]{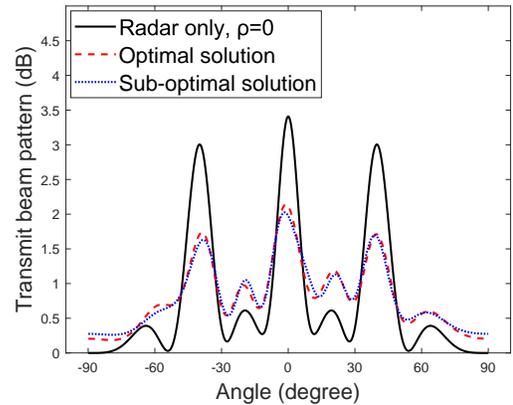} 
  }

  \caption{Radar beam pattern in different weight $\rho$: (a) $\rho=0.1$; (b) $\rho= 0.5$.} 
  \label{fig2} 
\end{figure}

In Fig.~\ref{fig2}, we evaluate the radar transmit beam patterns for the optimal beamforming scheme and sub-optimal beamforming scheme. The transmit beam patterns are depicted in Fig.~\ref{fig:subfig:near-field} for $\rho = 0.1$ and Fig.~\ref{fig:subfig:far-field} for $\rho = 0.5$, respectively. 
The radar only beam pattern is also provided as the optimal beam for radar sensing for comparison, which is obtained by solving \eqref{problem2} with $\rho=0$. From Fig.~\ref{fig2}, we can see that when the weighting factor is small, e.g., $\rho = 0.1$, the beam patterns of optimal and sub-optimal approach that of the only radar optimal beamforming, whereas the peak value of main beams of optimal scheme and sub-optimal scheme decreases significantly compared to that of only radar optimal beam pattern when $\rho = 0.5$. This illustrates the effect of WPT on radar sensing. Moreover, we notice that the average beam pattern of the sub-optimal scheme is comparable to that of the optimal scheme, which demonstrates the effectiveness of our proposed low-complexity sub-optimal design.

\begin{figure}
\centering
\includegraphics[scale=0.5]{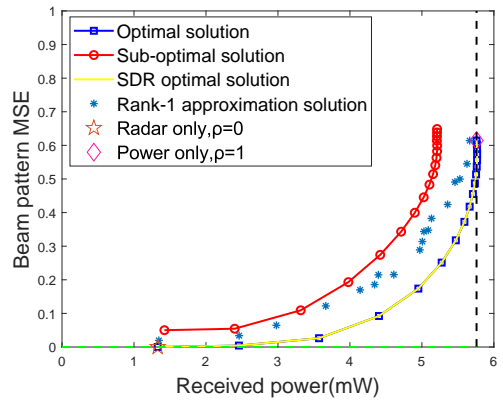}
\caption{Trade-off between radar beam pattern MSE and received power. } 
\label{fig3}
\end{figure}

In Fig.~\ref{fig3}, we explicitly demonstrate the trade-off between WPT and radar sensing for different beamforming schemes, where points "Radar only" and "WPT only" indicate the performance boundaries of radar and WPT, respectively. Meanwhile, we can observe that the radar beam pattern MSE of both optimal and sub-optimal designs increases with the increment of received power of users, which implies that the increase of received power is at the expense of radar performance. 
Moreover, we can see that the trade-off performance gap between the optimal scheme and sub-optimal scheme increases as the received power increases. This is because as the received power requirement increases, more transmit power is required to allocate to MRT beamforming that doesn't take the radar performance into account in sub-optimal scheme.
Besides, we also compare the performance of our proposed optimal solution with the commonly applied, and our proposed optimal solution is better than that of the rank-1 approximation solution in terms of beam pattern MSE, while achieve the same performance as the SDR optimal solution, this is consistent with our numerical results in Fig.~\ref{fig4}.


	\section{Conclusions}
	\label{sec:Conclusions}

In this letter, we studied a new ISWPT setup, in which the radar sensing and wireless power transfer are integrated into one hardware platform. We characterized the fundamental trade-off between radar sensing and wireless power transfer by optimizing transmitted beamforming vectors.  Both globally optimal solution and low-complexity sub-optimal solution are provided. Simulation results demonstrated the performance trade-off for integrated radar sensing and wireless power transmission, and also verified the effectiveness of our proposed beamforming designs.


\begin{appendices}

\section{Proof of Theorem \ref{thm:MRT}}
\label{Appendix:B}

Define ${\bf G}^{*} = \left( {\bf G}{\bf G}^{H} \right)^{-1}{\bf G}$. \eqref{MRTR} is then rewritten as
\begin{equation}
\label{MRTR2}
{\bf G}^{*} {\bf R} {\bf G}^{* H} = {\bf \Lambda}^2. 
\end{equation}

The eigenvalue decomposition of ${\bf R}$ is given by ${\bf R} ={\bf U}_{\rm r} {\bf \Lambda}_{\rm r} {\bf U}_{\rm r}^H$,  where ${\bf U}_{\rm r} \in \mathbb{C}^{N \times N}$ is the eigenvector matrix and ${\bf \Lambda}_{\rm r} \in \mathbb{C}^{N \times N}$ is the eigenvalue dialog matrix. The column vectors of ${\bf U}_{\rm r}$ form a set of $N$-dimensional orthogonal bases, thus there must exist ${\bf U}_{\rm r} = {\bf U}_{\rm r}^{*}$ that satisfy ${\bf G}^{*} {\bf U}_{\rm r}^{*} = \left[ {\bf I}_M,{\bf 0}_{M \times (N-M)} \right]$. Meanwhile, if we set ${\bf \Lambda}_{{\rm r}\left[ 1:M,1:M \right]} = {\bf \Lambda}^2$, 
then \eqref{MRTR2} will hold. Therefore, there always exists ${\bf R} = {\bf U}_{\rm r}^{*} {\bf \Lambda}_{\rm r} {\bf U}_{\rm r}^{*H} $ satisfies 
\begin{equation}
{\bf G} {\bf R } {\bf G}^{H} = {\bf G}{\bf G}^{H}{\bf \Lambda}^2 {\bf G}{\bf G}^{H}.
\end{equation}

Next we compute $\bf W$ from a given $\bf R$. Denote the Cholesky decomposition of $\bf R$ as ${\bf R} = {\bf D}{\bf D}^H$. Then, by performing the QR decomposition to $\left({\bf G}{\bf D}\right)^H$, we have ${\bf G}{\bf D} = \left[ {\bf L}, {\bf 0}_{M \times (N-M)} \right]{\bf U}$. Hence, we have
\begin{equation}
\label{MRTW}
{\bf G}{\bf D}{\bf D}^H{\bf G}^H = {\bf LL}^H = {\bf G} {\bf R } {\bf G}^{H}. 
\end{equation}

Similarly, by defining ${\bf H} = \left[ {\bf G}{\bf G}^{H}{\bf \Lambda},{\bf 0}_{M \times (N-M)} \right]$ and writing the QR decomposition of ${\bf H}^H$ as ${\bf H} = \left[ {\bf L}_{\rm H}, {\bf 0}_{M \times (N-M)} \right]{\bf U}_{\rm H}$, we have
\begin{equation}
\label{MRTW2}
{\bf H}{\bf H}^H = {\bf L}_{\rm H}{\bf L}^H_{\rm H} = {\bf G} {\bf R } {\bf G}^{H}.
\end{equation}

Comparing \eqref{MRTW} and \eqref{MRTW2}, we find that ${\bf L}={\bf L}_{\rm H}$. Thus, we can set 
\begin{equation}
{\bf W} = {\bf DU}^H{\bf U}_{\rm H},
\end{equation}
which satisfy
\begin{subequations}
\label{determen W}
\begin{align}
&~~~~~~{\bf W}{\bf W}^H = {\bf DU}^H{\bf U}_{\rm H}{\bf U}_{\rm H}^H{\bf U}{\bf D}^H = {\bf R}, \\  & {\bf G} {\bf W} = {\bf G}{\bf DU}^H{\bf U}_{\rm H} = \left[ {\bf G}{\bf G}^{H}{\bf \Lambda},{\bf 0}_{M \times (N-M)} \right].
\end{align}
\end{subequations}

Thus, the proof is completed.
\end{appendices}

	\bibliographystyle{IEEEtran}
	\bibliography{IEEEabrv,refs}

\end{document}